\documentclass[12pt,a4paper]{article}

\usepackage[utf8]{inputenc}
\usepackage[english]{babel}
\usepackage{fullpage}

\usepackage{amsmath}
\usepackage{amssymb}
\usepackage{amsthm}
\usepackage{fixmath}

\title{A recursive normalizing one-step reduction strategy \\
for the distributive lambda calculus}
\author{Anton Salikhmetov}

\newtheorem{theorem}{Proposition}
\newtheorem{definition}{Definition}

\begin{document}
\maketitle

\begin{abstract}
We positively answer the question A.1.6 in \cite{ustica}:
``Is there a recursive normalizing one-step reduction strategy
for micro $\lambda$-calculus?''
Micro $\lambda$-calculus refers to an implementation of the
$\lambda$-calculus due to \cite{revesz},
implementing $\beta$-reduction by means of ``micro steps''
recursively distributing a $\beta$-redex ${(\lambda x.M)\, N}$
over its body $M$.
\end{abstract}

\section{Inner spine strategy}

First, we provide ``micro $\lambda$-calculus'' with a more
systematic name\footnote{Vincent van Oostrom suggested
the following constructions a little differently in private
correspondence \cite{private}.
We took the liberty to change some notations hopefully
simplifying the proofs of the two core propositions.}.
\begin{definition}
\textit{Distributive reduction} is defined as $$
\beta_d = \beta_d^i \cup \beta_d^c \cup \beta_d^l \cup \beta_d^a,
$$ where
\begin{eqnarray*}
\beta_d^i &=& \bigl\{((\lambda x. x)\, M, M)
\bigm| M \in \Lambda\bigr\}, \\
\beta_d^c &=& \bigl\{((\lambda x. y)\, M, y)
\bigm| M \in \Lambda\bigr\}, \\
\beta_d^l &=& \bigl\{((\lambda x. \lambda y. M)\, N,
\lambda y. (\lambda x. M)\, N)
\bigm| M, N \in \Lambda\bigr\}, \\
\beta_d^a &=& \bigl\{((\lambda x. M\, N)\, P,
(\lambda x. M)\, P\, ((\lambda x. N)\, P))
\bigm| M, N, P \in \Lambda\bigr\}.
\end{eqnarray*}
Additionally, we denote the following binary relations:
\begin{eqnarray*}
(M, N) \in \beta_d  &\Rightarrow& \forall C[\phantom M]: C[M] \rightarrow_d C[N], \\
(M, N) \in \beta_d^i &\Rightarrow& \forall C[\phantom M]: C[M] \rightarrow_i C[N], \\
(M, N) \in \beta_d^c &\Rightarrow& \forall C[\phantom M]: C[M] \rightarrow_c C[N], \\
(M, N) \in \beta_d^l &\Rightarrow& \forall C[\phantom M]: C[M] \rightarrow_l C[N], \\
(M, N) \in \beta_d^a &\Rightarrow& \forall C[\phantom M]: C[M] \rightarrow_a C[N].
\end{eqnarray*}
\end{definition}
\begin{theorem}
Any ${\beta_d}$-redex is a $\beta$-redex, and vice versa.
\end{theorem}
\begin{proof}
The proposition directly follows from the definition of
${\beta_d}$.
\end{proof}

Normalisation of our strategy answering the previously open
question, relies on the one hand on normalization of spine
reductions for the ordinary $\lambda$-calculus, and on the other
hand on termination of pure distribution steps, as encountered
in the $\lambda$-calculus with explicit substitutions
${\lambda x}$.
\begin{definition}
\textit{Inner spine strategy} contracts the innermost redex among
spine redexes (see the definition 4.7 in \cite{needred}).
\end{definition}
\subsection{Correctness of distributive reduction}

A term is a distributive redex if and only if it is a
$\beta$-redex, hence distributive and $\beta$-normal forms
coincide.
In turn, the spine redexes with respect to distributive reduction
coincide with those for ordinary $\beta$-reduction.
If $M$ distributively rewrites to ${M'}$, then in general $M$
need not $\beta$-rewrite to ${M'}$, but $M$ and ${M'}$ are
$\beta$-convertible.
\begin{theorem}
${M \rightarrow_d N \Rightarrow M =_\beta N}$.
\end{theorem}
\begin{proof}
Let us consider each subset of ${\beta_d}$.
\begin{enumerate}
\item If ${M \rightarrow_i N}$, then for some $P$ $$
M \equiv (\lambda x. x)\, P \land N \equiv P,
$$ but then $$
(\lambda x. x)\, P \rightarrow_\beta x[x := P] \equiv P.
$$
\item If ${M \rightarrow_c N}$, then for some $P$ $$
M \equiv (\lambda x. y)\, P \land N \equiv y,
$$ but then $$
(\lambda x. y)\, P \rightarrow_\beta y[x := P] \equiv y.
$$
\item If ${M \rightarrow_l N}$, then for some $P$, $Q$ $$
M \equiv (\lambda x. \lambda y. P)\, Q \land
N \equiv \lambda y. (\lambda x. P)\, Q,
$$ but then $$
(\lambda x. \lambda y. P)\, Q \rightarrow_\beta
(\lambda y. P)[x := Q] \equiv \lambda y. P[x := Q]
\leftarrow_\beta \lambda y. (\lambda x. P)\, Q.
$$
\item If ${M \rightarrow_a N}$, then for some $P$, $Q$, $R$ $$
M \equiv (\lambda x. P\, Q)\, R \land
N \equiv (\lambda x. P)\, R\, ((\lambda x. Q)\, R),
$$ but then
\begin{eqnarray*}
(\lambda x. P\, Q)\, R \rightarrow_\beta (P\, Q)[x := R] \equiv\\
P[x := R]\, Q[x := R] \leftarrow_\beta
(\lambda x. P)\, R\, ((\lambda x. Q)\, R),
\end{eqnarray*}
\end{enumerate}
Since we have traversed ${\beta_d^i}$, ${\beta_d^c}$,
${\beta_d^l}$, ${\beta_d^a}$, the proposition also stands for
${\beta_d}$.
\end{proof}

\subsection{Useful definitions}

\begin{definition}
\textit{Full $\beta$-development ${M^\bullet}$} of a term $M$ is
the term obtained by $\beta$-contracting all redexes of $M$.
\end{definition}
\begin{definition}
A step is called \textit{destructive} if the redex contracted is
of shape
$$
((\lambda x. (\lambda y. P)\, Q))\, R,
$$
that is, in case of distribution of $N$ over an application
${(\lambda y. M_1)\, M_2}$ which itself is a redex.
\end{definition}

Our strategy relies on the observation that distributive
reduction is preserved when projecting every term to its full
$\beta$-development, as long as the steps of the former are not
$\beta$-destructive.
Non-destructive steps will be mapped to $\beta$-reduction
sequences by $\bullet$.

Instead of proving this general fact, we note inner spine steps
are non-destructive by innerness, and show that each such inner
spine step is mapped to at most a signle $\beta$-reduction step
by $\bullet$.
Moreover, in case a distributive inner spine step is mapped to
an empty step by $\bullet$, i.~e. if it is erased, then that step
did not create a redex, hence it is a purely distributive step.
This can be expressed formally by mapping the step to an $x$-step
in Bloo and Rose's $\lambda$-calculus with explicit substitutions
${\lambda x}$ \cite{bloo}, via the following operation.
\begin{definition}
\textit{Explicification ${M^\diamond}$} of a term $M$ is obtained
by replacing each of its redexes ${(\lambda x. P)\, Q}$ by the redex
${P \langle x := Q \rangle}$ in the $\lambda$-calculus with
explicit substitutions ${\lambda x}$.
\end{definition}

\subsection{Proof of normalizing property}

\begin{theorem}
If ${M \rightarrow_d N}$, then either the step
${M^\bullet \rightarrow_\beta N^\bullet}$ contracts a spine
redex, or ${M^\bullet \equiv N^\bullet}$ and
${M^\diamond \rightarrow_x N^\diamond}$.
\end{theorem}
\begin{proof}
Let us consider each of the possible cases.
\begin{enumerate}
\item If an inner spine step ${M \rightarrow N}$ is due to
${M \rightarrow_d N}$, then ${M^\bullet \equiv N^\bullet}$ and
${M^\diamond \rightarrow_x N^\diamond}$ immediately follow from
the proposition about correctness of distributive reduction.
\item If the innner spine step ${M\, P \rightarrow N\, P}$
is due to ${M \rightarrow_d N}$, then relying on (1) having been
proved, let us consider the following three possible options.
\begin{enumerate}
\item ${M\, P}$ is a $\beta$-redex.
Then ${M \equiv \lambda x. M'}$ and ${N \equiv \lambda x. N'}$
for some ${M'}$ and ${N'}$, and ${M' \rightarrow_\beta N'}$,
hence either $$
(M\, P)^\bullet \equiv M'^\bullet[x := P^\bullet]
\rightarrow_\beta
N'^\bullet[x := P^\bullet] \equiv (N\, P)^\bullet
$$ is a spine step, or $$
(M\, P)^\bullet \equiv M'^\bullet[x := P^\bullet] \equiv
N'^\bullet[x := P^\bullet] \equiv (N\, P)^\bullet
$$ and $$
(M\, P)^\diamond \equiv
M'^\diamond \langle x := P^\diamond \rangle \rightarrow_x
N'^\diamond \langle x := P^\diamond \rangle \equiv
(N\, P)^\diamond.
$$
\item ${M\, P}$ is not, but ${N\, P}$ is a $\beta$-redex.
Then ${N \equiv \lambda x. N'}$ for some ${N'}$, and either
${M \rightarrow_i N}$ and ${M \equiv (\lambda x. x)\, N}$,
or ${M \rightarrow_l N}$ and for some ${M'}$, ${N''}$
\begin{eqnarray*}
M &\equiv& (\lambda x y. M')\, N''; \\
N' &\equiv& (\lambda x. M')\, N''.
\end{eqnarray*}
The case of ${\beta_d^i}$ is trivial, while for ${\beta_d^l}$
we have
\begin{eqnarray*}
(M\, P)^\bullet \equiv
((\lambda x y. M')\, N'')^\bullet\, P^\bullet \equiv \\
(\lambda y. M')^\bullet[x := N''^\bullet]\, P^\bullet \equiv
(\lambda y. M'^\bullet[x := N''^\bullet])\, P^\bullet
\end{eqnarray*}
and
\begin{eqnarray*}
(N\, P)^\bullet \equiv ((\lambda y. N')\, P)^\bullet \equiv \\
(\lambda y. (\lambda x. M')\, N'')\, P^\bullet \equiv
M'^\bullet[x := N''^\bullet][y := P^\bullet],
\end{eqnarray*}
then ${(M\, P)^\bullet \rightarrow_\beta (N\, P)^\bullet}$,
hence the proposition stands since a head redex is a spine redex.
\item Neither of ${M\, P}$ and ${N\, P}$ is a $\beta$-redex, then ${(N\, P)^\bullet \equiv N^\bullet\, P^\bullet}$, ${(N\, P)^\diamond \equiv N^\diamond\, P^\diamond}$, and
the proposition stands.
\end{enumerate}
\item If an inner spine step ${P\, M \rightarrow P\, N}$ is due
to ${M \rightarrow_d N}$, then relying on (1) having been proved
let us note that ${P\, M}$ cannot be a redex.
Therefore, ${P\, N}$ is not a redex either.
But then again either $$
(P\, M)^\bullet \equiv P^\bullet\, M^\bullet \rightarrow_\beta
P^\bullet\, N^\bullet \equiv (P\, N)^\bullet
$$ is a spine step, or $$
(P\, M)^\bullet \equiv P^\bullet\, M^\bullet \equiv
P^\bullet\, N^\bullet \equiv (P\, N)^\bullet
$$ and $$
(P\, M)^\diamond \equiv P^\diamond\, M^\diamond \rightarrow_x
P^\diamond\, N^\diamond \equiv (P\, N)^\diamond.
$$

\item If an inner spine step
${\lambda x. M \rightarrow \lambda x. N}$ is due to
${M \rightarrow_d N}$, then relying on (1) having been proved
we immediately get that either $$
(\lambda x. M)^\bullet \equiv \lambda x. M^\bullet
\rightarrow_\beta
\lambda x. N^\bullet \equiv (\lambda x. N)^\bullet
$$ is a spine step, or $$
(\lambda x. M)^\bullet \equiv \lambda x. M^\bullet \equiv
\lambda x. N^\bullet \equiv (\lambda x. N)^\bullet
$$ and $$
(\lambda x. M)^\diamond \equiv \lambda x. M^\diamond
\rightarrow_x
\lambda x. N^\diamond \equiv (\lambda x. N)^\diamond.
$$
\end{enumerate}
Thereby, the possible cases have been treated thoroughly.
\end{proof}
\begin{theorem}
Inner spine strategy is normalizing.
\end{theorem}
\begin{proof}
By the previous proposition, an infinite distributive reduction
from some term $M$ having a normal form $\hat M$, would give rise
to an infinite spine $\beta$-reduction from ${M^\bullet}$, unless
from some moment $N$ on in the distributive reduction all further
terms are mapped to ${N^\bullet}$.
But then by the same proposition, the infinite distributive
reduction from $N$ would give rise to an infinite $x$-reduction
from ${N^\diamond}$.

Infinite spine $\beta$-reductions are impossible from
${M^\bullet}$ since $M$ and ${M^\bullet}$ are
$\beta$-convertible, hence have the same $\beta$-normal form
${\hat M}$, and spine strategies are needed strategies, hence
normalising \cite{needred}.

In turn, infinite $x$-reductions are impossible since
$x$-reduction (the substitution rules) is known to be terminating
for the ${\lambda x}$-calculus \cite{bloo}.
\end{proof}
The essence of our strategy is to avoid destruction of redexes.
In particular, the inner spine strategy avoids
that distribution of the outer redex in
${(\lambda x.(\lambda y.P)\, Q)\, R}$ destroys the inner one,
thereby blocking Klop's counterexample to preservation of strong
normalisation for distributive reduction.

\end{document}